# Low temperature MQ NMR dynamics in dipolar ordered state

G. B. Furman [1,2], V. M. Meerovich [1], V. L. Sokolovsky [1]

[1] Department of Physics, Ben Gurion University, Beer Sheva 84105, Israel

[2] Zefat College, Zefat, Israel

We investigate analytically and numerically the Multiple Quantum (MQ) NMR dynamics in dipolar ordered spin systems of nuclear spins $1/2$ at a low temperatures. We suggest two different methods of MQ NMR. One of them is based on the measurement of the dipolar energy. The other method uses an additional resonance $(\pi/4)_y$ -pulse after the preparation period of the standard MQ NMR experiment in solids and allows one to measure the Zeeman energy. The both considered methods are sensitive to the contribution of remote spins in the interaction and to the spin system structure. The QS method is sensitive to the spin number in the molecule while the PS method gives very similar time dependecies of the intensities of MQ coherences for different spin numbers. It is shown that the use of the dipolar ordered initial state has the advantage of exciting the highest order MQ coherences in clusters of $4m$ identical spins, where $m = 1, 2, 3...$, that is impossible to do with the standard MQ method. MQ NMR methods based on the dipolar ordered initial states at low temperatures complement the standard NMR spectroscopy for better studying structures and dynamic processes in solids.

## Introduction

Multiple-quantum (MQ) NMR spectroscopy has been a useful tool in fundamental studies of physics and chemistry. It has been used to probe spin dynamics and structure of solids, counting the number of spins in impurity clusters [1, 2] and to analyze NMR spectra [3].

Many current studies [4-11] focus on how to harness the potential of MQ NMR in the development of quantum computation and quantum metrology. The growth in interest in MQ NMR applications stimulates intensive theoretical research in the field of MQ NMR. However, most of the studies deal with high temperature region [12-14], while at low temperature MQ NMR spectroscopy possesses specific properties [15, 16]. For example, at low temperature the spin-lattice relaxation processes can be disregard during MQ experiments and, as a consequence, can be used to obtain accurate and precise information about molecular dynamic processes and molecular structure. It is shown that the growth of intensities of MQ coherence gets faster when the temperature decreases, and the intensities of multiple quantum coherences can be negative at low temperatures [16]. A few years later, it was proposed to use a so-called dipolar ordered state, a state in which nuclear spins are oriented along an internal local magnetic field [17], as the initial state in high temperature MQ NMR experiments [18] to faster form the correlations in spin systems [19, 20]. As a matter of fact, it was found [19, 20] that in experiments with the dipolar ordered state at high temperature the intensities of MQ coherences were several times higher than the analogous coherence in the standard MQ NMR experiment in which nuclear spins were oriented along an external magnetic field.

Consideration of the MQ dynamics at low temperatures is important from several points of view. Firstly, MQ coherences are created several times faster and thus attain larger intensity values than at higher temperatures. Second, the influence of various decoherence processes at low temperatures is lower. Third, the developed MQ coherence methods can be successfully applied to creation of entangled states which, in turn, are widely used in quantum technology.

In the present paper, we demonstrate the application of MQ NMR spectroscopy in dipolar ordered spin systems at low temperatures.

## Low temperature dipolar ordered state

Let us start with consideration of a system of nuclear spins ($s=1/2$) coupled by the dipole-dipole interaction (DDI) in a strong external magnetic field, $\vec{H}_0$ directed along the $z$-axis. The effect of the dipolar interaction is small in comparison with the Zeeman interaction and, in the first-order perturbation, the secular part of the DDI Hamiltonian plays the main role in a spin dynamics. This secular part has the following form [17]

$$\mathbf{H}_{dz} = \sum_{j<k} D_{jk} \left[ I_{jz}I_{kz} - \frac{1}{4}\left(I_j^+ I_k^- + I_j^- I_k^+\right)\right], \quad (1)$$

where $D_{jk} = \frac{\gamma^2 \hbar}{r_{jk}^3}(1-3\cos^2\theta_{jk})$ is the coupling constant between spins $j$ and $k$, $\gamma$ is the gyromagnetic ratio, $r_{jk}$ is the distance between spins $j$ and $k$, $\theta_{jk}$ is the angle between the internuclear vector $\vec{r}_{jk}$ and the external magnetic field $\vec{H}_0$. $I_{j\alpha}(\alpha=x,y,z)$ is the projection of the angular spin momentum operator on the axis $\alpha$; $I_j^+$ and $I_j^-$ are the raising and lowering operators of spin $j$, $j=1,...,N$, $N$ is the number of spins in the system, $\hbar$ is the Plank constant.

Schematically, the standard MQ NMR experiments can be presented as four stages taking place at successive time periods: preparation, evolution, mixing and detection [1]. MQ coherences are created by a multipulse sequence consisting of eight-pulse cycles during the preparation period [1]. In the rotating frame [17], the average Hamiltonian describing the MQ dynamics during the preparation period can be written as [1]

$$\mathbf{H}_{MQ} = \mathbf{H}^{(2)} + \mathbf{H}^{(-2)}, \quad (2)$$

where $\mathbf{H}^{(\pm 2)} = -\frac{1}{4}\sum_{j<k} D_{jk} I_j^\pm I_k^\pm$.

The density matrix of the spin system, $\rho(\tau)$, at the end of the preparation period with duration $\tau$ is

$$\rho(\tau) = U(\tau)\rho(0)U^+(\tau), \quad (3)$$

where $U(\tau) = \exp(-i\tau(\mathbf{H}^{(2)} + \mathbf{H}^{(-2)}))$ is the evolution operator within the preparation period and $\rho(0)$ is the initial density matrix of the system.

Here we consider MQ NMR dynamics with the initial dipolar ordered state of the spin system when the Hamiltonian of the system is determined by Eq. (1). The dipolar ordered state at a low temperature can be obtained using the method of adiabatic demagnetization in rotating frame [17]. Under the thermal equilibrium, the system in the dipolar ordered state can be described as

$$\rho(0) = \frac{1}{Z}\exp(-\beta\mathbf{H}_{dz}), \qquad (4)$$

where $\beta$ is the inverse spin temperature, $Z = Tr\{\exp(-\beta\mathbf{H}_{dz})\}$ is the partition function. Below we will calculate the intensities of the MQ coherences at low temperatures using two different approaches [19, 20]. The first approach is based on the calculation of the MQ coherences simulating the experimental method in which the measurable value is dipolar signal [21]. The dipolar signal corresponds to the observed signal in phase with the exciting radio-frequency pulse (phase signal ($PS$) method) [17]. As the second approach we will consider MQ NMR method which consists in applying an additional resonance $\frac{\pi}{4}$ -pulse after the preparation period of the standard MQ NMR experiment [19]. The signal detected in quadrature with the radio-frequency pulse is proportional to the Zeeman energy (quadrature signal ($QS$) method) [17].

## MQ NMR dynamics with dipolar ordered initial state
*PS method*

Starting with the initial dipolar ordered condition (4), the density matrix, $\rho(\tau,t)$, after the three stages realized similarly to the standard MQ NMR experiment [1, 2] can be written as

$$\rho(\tau,t) = U^+(\tau)e^{-i\delta I_z}\rho(\tau)e^{i\delta I_z}U(\tau), \qquad (5)$$

where $\rho(\tau)$ is the density matrix at the end of the preparation period according to Eq. (3), $\delta$ is the frequency offset in the evolution period of the duration $t$. The offset is a result of applying the time proportional phase incrementation (TPPI) method [1]. In Eq. (5) the unitary transformation with operator $U(\tau)$ describes the mixing period with duration $\tau$. It should be emphasized that in Eq. (5) the initial density matrix defined by Eq. (4) over the whole temperatures range in contrast to early published papers [19, 20] where MQ NMR spectra was analyzed in the high temperature approximation.

Using Eqs. (5) and (4) the dipolar signal, $S_D(t)$, after the three stages can be represented as

$$S_D(\tau,t) = Tr\{e^{-i\delta I_z t}\rho(\tau)e^{i\delta I_z t}\rho_D(\tau)\} \qquad (6)$$

where

$$\rho_D(\tau) = U(\tau)H_{dz}U^+(\tau). \qquad (7)$$

We expand the density matrix, $\rho(\tau)$, given by (3), as

$$\rho(\tau) = \sum_n \rho_n(\tau), \qquad (8)$$

where the term $\rho_n(\tau)$ is responsible for the MQ coherence of the $n$-th order. Following [20] we expand the operator, $\rho_D(\tau)$, given by (7), as

$$\rho_D(\tau) = \sum_n \rho_{-n}^D(\tau). \qquad (9)$$

Using Eqs. (7) - (9) one can rewrite the expression for the observable dipolar signal (6) in terms of the intensities of MQ coherences [19, 20]

$$S_D(\tau,t) = \sum_n e^{-in\delta t} J_{nD}(\tau) \tag{10}$$

where $J_{nD}(\tau)$ is the intensity of MQ coherency of $n$-th order as a function of $\tau$ [19, 20]

$$J_{nD}(\tau) = Tr\{\rho_n(\tau)\rho_{-n}^D(\tau)\}. \tag{11}$$

*QS method*
In the previous section we showed that starting from the initial condition (4) the standard procedure of MQ NMR experiments leads to observable signal in phase with the exciting radio-frequency pulse. In order to detect the signal in quadrature with the radio-frequency pulse we introduce a $\phi_y$-pulse turning spins around the $y$-axis by the angle $\phi$ after the preparation period. Following [19] for with the initial condition (4), after the mixing period the quadrature signal, $S_z(t)$, can be defined as

$$S_z(\tau,t) = Tr\{\rho(\tau,t)I_z\} \tag{12}$$

where $\rho(\tau,t)$ is the density matrix after the three stages with the modified MQ NMR pulse sequence [19]

$$\rho(\tau,t) = U^+(\tau)e^{-i\mathcal{H}_z t}\rho_\phi(\tau)e^{i\mathcal{H}_z t}U(\tau). \tag{13}$$

and $\rho_\phi(\tau)$ is the density matrix just after the $\phi_y$-pulse [19]

$$\rho_\phi(\tau) = e^{-i\phi_y I_y}U(\tau)\rho(0)U^+(\tau)e^{i\phi_y I_y}. \tag{14}$$

Using Eqs. (13) - (14) and the initial condition (4) the quadrature signal (12) can be rewritten [19]

$$S_z(\tau,t) = Tr\{e^{-i\mathcal{H}_z t}\rho_\phi(\tau)e^{i\mathcal{H}_z t}\rho_z(\tau)\} \tag{15}$$

where the operator [22]

$$\rho_z(\tau) = U(\tau)I_z U^+(\tau), \tag{16}$$

coincides with the density matrix at the end of the preparation period of the standard MQ NMR experiment with the thermodynamical equilibrium density matrix as the initial condition [1]. The density matrix $\rho_z(\tau)$ can be represented in the following form [22]

$$\rho_z(\tau) = \sum_n \rho_n^z(\tau) \tag{17}$$

where the term $\rho_n^z(\tau)$ is responsible for the MQ coherence of the $n$-th order. Using Eqs. (16), (17) and $\phi_y = \frac{\pi}{4}$ one can rewrite the expression for the observable signal in terms of the intensities of MQ coherences [19]

$$S_z(t) = -i\sum_n e^{-in\delta t} J_{nz}(\tau) = i\sum_n e^{in\delta t} J_{nz}^*(\tau) = i\sum_n e^{-in\delta t} J_{-nz}^*(\tau), \tag{18}$$

where [19]

$$J_{nz}(\tau) = iTr\{\rho_\phi(\tau)\rho_{nz}^z(\tau)\} \tag{19}$$

and the quadrature signal $S_z(t)$ is always real.

# Analysis of the time evolution of MQ coherences

We will first obtain an exact solution for a four-spin system with the nearest-neighbor interactions starting with the initial condition (4) using both considered MQ NMR methods, $PS$ and $QS$. Then taking into account the dipolar interaction of all spins we will numerically investigate the MQ NMR dynamics with the initial condition (4) considering four- and six-spin circles and four- and eigth-spin linear chains and ten-spin system which simulate real spin structures. Examples of such structures are dipolar-coupled proton spins of benzene molecule $C_6H_6$ (circle of six spins) and xenon tetrafluoride with chemical formula $XeF_4$ (its crystalline square planar structure was determined by both NMR spectroscopy and by neutron diffraction studies [23, 24]), and calcium hydroxyapatite $Ca_5(OH)(PO_4)_3$ and calcium fluorapatite $Ca_5F(PO_4)_3$ (proton and fluorine chains) [25]. As a simulated spin system consisting of ten spins we consider cyclopentane molecules, $C_5H_{10}$ with ten hydrogen atoms [26].

*MQ dynamics in a four-spin system with nearest-neighbor interactions*

To describe a one-dimensional solids, a quantum spin chain model can be applied with the fixed angle $\theta_{jk}$ for any pairs of spins along the chain, and the lattice spacing, or the nearest-neighbor distance, $r_{j,j+1}$, is also a constant.

For a chain of 4 spins, we obtain an analytical solution describing the dependence of the zeroth, second and fourth order intensities on the temperature and time duration of the preparation period, $\tau$. The analytical calculation is performed using the software based on the MATEMATICA package. Previously, the exact solutions were obtained with the standard initial condition in approximations of high- and low temperatures [14, 15] and it has been shown that only the zeroth- and second-order quantum coherences appears in MQ NMR spectra. The uniqueness of the present solution is non-zero value of intensity of the fourth-order coherence for both, $PS$ and $QS$, methods. Here we give only an expression describing this coherence as a function of the dimensionless time $\bar{t} = D\tau$ and inverse spin temperature $\bar{\beta} = \frac{D}{kT}$ (here $D$ is the dipolar coupling constant between the nearest neighbors ):

$$J_{4D} = \frac{e^{-\frac{3\bar{\beta}}{4}}\left(-2 + 5\cos 2t - 3\cos 2t\sqrt{5}\right)A}{40\left(\frac{3}{\sqrt{2}}\right)Z_4} \tag{20}$$

for $PS$ method and

$$J_{4z} = \frac{e^{-\frac{3\bar{\beta}}{4}} AB}{160 Z_4 \left(\sqrt{5}-3\right)\sqrt{3+\sqrt{5}}} \tag{21}$$

for $QS$ method, respectively. In Eqs. (20) and (21) $Z_4$ is the partial function

$$Z_4 = \left(1 + 2e^{-\frac{3\bar{\beta}}{4}} + e^{\frac{\bar{\beta}}{4}} + 2e^{\frac{\bar{\beta}}{4}}\cosh\frac{\bar{\beta}\sqrt{5}}{4} + 4e^{-\frac{\bar{\beta}}{8}}\cosh\frac{\bar{\beta}\sqrt{5}}{8} + 4e^{-\frac{\bar{\beta}}{8}}\cosh\frac{\bar{\beta}\sqrt{13}}{8} + 2e^{\frac{3\bar{\beta}}{8}}\cosh\frac{\bar{\beta}\sqrt{33}}{8}\right)$$

(22)

and the quantities $A$ and $B$ take the following forms

$$A = \left(\cos^2 \bar{t} - \cos^2 \sqrt{5}\bar{t} + e^{\bar{\beta}}\sin^2 \bar{t} - e^{\frac{9}{8}\bar{\beta}}\sin^2 \sqrt{5}\bar{t}\left(\cosh\frac{\bar{\beta}\sqrt{33}}{8} - \frac{7\sqrt{33}}{55}\sinh\frac{\bar{\beta}\sqrt{33}}{8}\right)\right) \tag{23}$$

and

$$B = (5+3\sqrt{5})\sin(\sqrt{6+2\sqrt{5}})\bar{t} + (15-7\sqrt{5})\sin(\sqrt{6-2\sqrt{5}})\bar{t}$$
$$-3(3\sqrt{5}-5)\sqrt{6+2\sqrt{5}}\cos\bar{t}\sin(\sqrt{5}\bar{t}) \tag{24}$$

In the high temperature approximation ($\bar{\beta} \ll 1$), the Eqs. (20) and (21) take the following forms

$$J_{4D} = -\frac{\sqrt{2}\bar{\beta}(3\cos 2\sqrt{5}\bar{t} - 5\cos 2\bar{t} + 2)}{76800}\left(20\cos 2\bar{t} - \frac{45}{2}\cos 2\sqrt{5}\bar{t} - 21\sqrt{5}\bar{t}\sin 2\sqrt{5}\bar{t} + \frac{5}{2}\right) \tag{25}$$

for $PS$ method, and

$$J_{4z} = -\frac{\bar{\beta}\left(-\frac{9}{128}\sin^2\sqrt{5}\bar{t} + \frac{21}{320}\sqrt{5}\bar{t}\cos\sqrt{5}\bar{t}\sin\sqrt{5}\bar{t} + \frac{1}{16}\sin^2\bar{t}\right)B}{160(\sqrt{5}-3)\sqrt{\sqrt{5}+3}} \tag{26}$$

for $QS$ method, respectively.

The intensities of MQ coherences of the fourth order as functions of dimensionless time $\bar{t}$ and inverse temperature $\bar{\beta}$, calculated according to Eqs. (20) and (21), are presented in Fig 1. MQ coherence intensities given by both methods are time periodic and increases with $\bar{\beta}$ (temperature decrease) up to the limits

$$J_{4D} = \frac{\left(1 - \frac{7\sqrt{33}}{55}\right)\sin^2(\sqrt{5}\bar{t})}{120}\left(2 - 5\cos 2\bar{t} + 3\cos(2\sqrt{5}\bar{t})\right) \tag{27}$$

and

$$J_{4z} = \frac{\left(1 - \frac{7\sqrt{33}}{55}\right)\sin^2(\sqrt{5}\bar{t})B}{320(3-\sqrt{5})\sqrt{3+\sqrt{5}}} \tag{28}$$

and reach the maximum $J_{4D} = |0.011|$ and $J_{4z} = |0.007|$. At low temperatures the intensities are few orders greater in magnitude than those at high temperatures.

Note that the standard MQ NMR method does not allow exciting the highest-order (fourth-order) multiple quantum (HOMQ) coherence for a four-spin system at any temperature [1].

*Numerical analysis of the time evolution of MQ coherences in dipolar coupling spin system*

Numerical calculations are performed using the dipolar coupling constants of $j$-th and $k$-th spins determined by $D/|j-k|^3$ and $D\left(\frac{\sin(\pi/N)}{\sin(\pi(j-k)/N)}\right)^3$ for a chain and a circle, respectively. The molecule cyclopentan has two pentagon circle and the reduced dipolar coupling constants, $\bar{D}_{jk} = D_{jk}/D_{11'}$, are following: $\bar{D}_{11} = 1$ (by the definition), $\bar{D}_{12} = -0.178$, $\bar{D}_{12'} = -0.002$, $\bar{D}_{13} = -0.093$, and $\bar{D}_{13'} = -0.26$ [26]. Here the indexes without touch note the coupling constants between spins placed in the same circle; the indexes with touch - the constants between spins in different circles. Other constants are determined from symmetry. In order to compare the results of the numerical simulations with the analogous ones at high temperatures, we introduce the normalized intensities of MQ coherences:

$J_{nD}(\tau)/\{Tr(\mathbf{H}_{dz}^2)\}^{1/2}$ and $J_{nz}(\tau)/\{Tr(I_z^2)\}^{1/2}$ for the $PS$ and $QS$ methods, respectively.

The dependence of the intensities of MQ coherences on the dimensionless time $\bar{t}=D\tau$ in a four-spin chain is presented in Fig. 2 for both, $PS$ and $QS$, methods and at low (Fig. 2a) and high (Fig. 2b) temperatures. It is evident that the suggested methods at low temperature can be considered as useful additions to the MQ NMR method at high temperatures. One can see (Fig. 2) that the low temperature methods are more preferable for experimental realizations, because they yield the signal of MQ coherence of the forth order which several times higher than the analogous coherence signal in the high temperature MQ NMR experiment.

Comparison of the low temperature methods of MQ NMR of the systems in the dipolar ordered state in circles consisting of four and six spins is given in Fig. 3 for MQ coherences of the fourth order. The $PS$ method gives the intensity about two orders of magnitude larger than that given by the $QS$ method. On the other hand, the $QS$ method is more sensitive to the spin system structure.

Fig. 4 shows the evolution of the normalized intensities of $8Q$ - coherences in the in eight- (Fig. 4a) and in the ten- (Fig. 4b) spin systems for both, $PS$ and $QS$, methods. Fig. 4 demonstrates the dependence of the intensity on the spin system structure: the signal of MQ coherence of the eighth order in the eight-spin system is about four orders of magnitude higher than the analogous intensity in the ten-spin system (in the cyclopentan).

Numerical calculations show that, using the dipolar ordered initial state, it is possible to excite HOMQ coherences in spin systems, in which the standard MQ NMR method does not work. For example, starting from the initial condition (4), the $4Q$ - and $8Q$ - coherences can be excited in four-and eight-spin systems, respectively.

Calculations show (see Figs. 1 - 4 and Eqs. (20) - (26)) that the intensities of MQ coherences can be negative in contrast to the standard MQ NMR experiments at high temperatures [1, 22, 27, 28]. It was shown [16] that a negative intensity appears also in the standard NMR experiments at low temperatures. This is due to the fact that the longitudinal magnetization is modulated by RF pulses and different frequency components have different signs [16].

## Conclusions

Two MQ NMR methods for the detection of MQ coherences starting from the dipolar ordered state at a low temperature have been considered. The first method, $PS$, is based on the measurement of the dipolar energy. This energy is a source of the information about MQ NMR dynamics. In order to observe a non-zero signal of MQ coherences using the second method, $QS$, an additional resonance $(\pi/4)_y$ -pulse should be applied after the preparation period of the standard MQ NMR experiment in solids, as was proposed early at high temperatures [19].

The both considered methods are sensitive to the type of the the dipole-dipole interaction (Fig. 2) and to the spin system structure (Fig. 4). The $QS$ method is sensitive to the spin number in the molecule while the $PS$ method gives very similar time dependecies of the coherence intensities of MQ coherences for any number of spins (Fig. 4).

It was shown that the use of the dipolar ordered initial state has the advantage: the HOMQ coherence can be excited in clusters of $4m$ identical spins, where $m=1,2,3...$, that impossible to do with the standard MQ method [1].

Therefore, MQ NMR methods based on the dipolar ordered initial states at low temperatures complement the standard NMR spectroscopy to better study structures and dynamic processes in solids.

## Acknowledgements

We are grateful to Prof. E.B. Fel'dman for useful discussions.

Figures

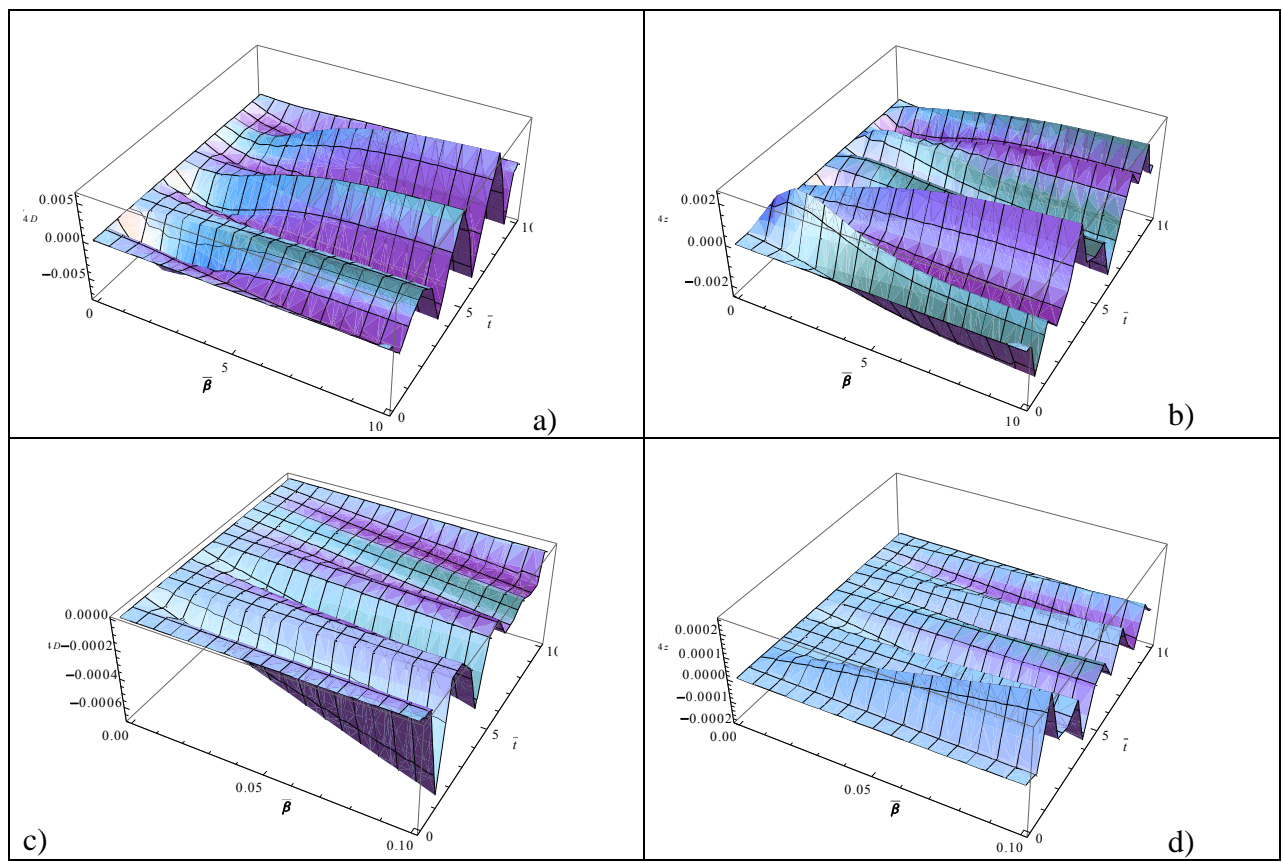

Fig. 1 The intensities of the fourth order MQ coherences in a linear chain of four spins coupled by the nearest-neighbor dipolar interaction as a function of the time and inverse temperature, calculated using Eqs. (20) and (21) : (a) $PS$ method and low temperature $\bar\beta > 1$ ; (b) $QS$ method and low temperature $\bar\beta > 1$ ; (c) $PS$ method and high temperatures $\bar\beta \ll 1$ ; (d) $QS$ method and high temperatures $\bar\beta \ll 1$ ;. Here and in other figures the indexes "$D$" and "$z$" note the results obtained by the $PS$ and $QS$ methods, respectively.

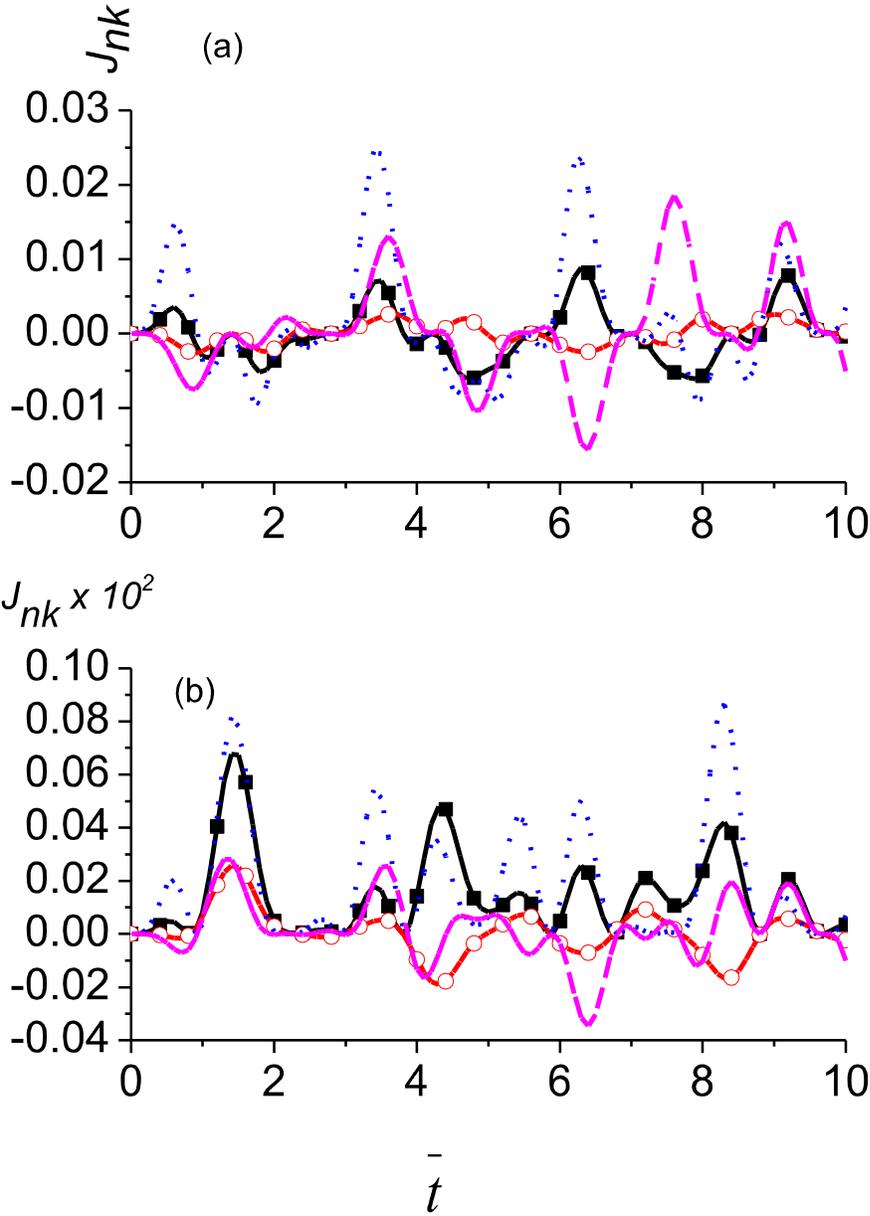

Fig. 2 The time dependence of the intensities of the MQ coherences of the fourth order in a linear chain of four spins coupled by the DDI for low, $\beta = 10$, (a) and high, $\beta = 0.1$, (b) temperatures. The curves present results of numerical calculations: black solid and red dashed - $PS$ and $QS$ methods for systems with nearest-neighbor interaction; blue dotted and pink dash-dotted - $PS$ and $QS$ methods taking into account dipolar interaction of all the spins, respectively. Black squares and open circles present calculations using Eqs. (20) and (21) simulating $PS$ and $QS$ methods for systems with nearest-neighbor interaction, respectively. The index "$k$" equals "$D$" and "$z$" for the results obtained by the $PS$ and $QS$ methods, respectively.

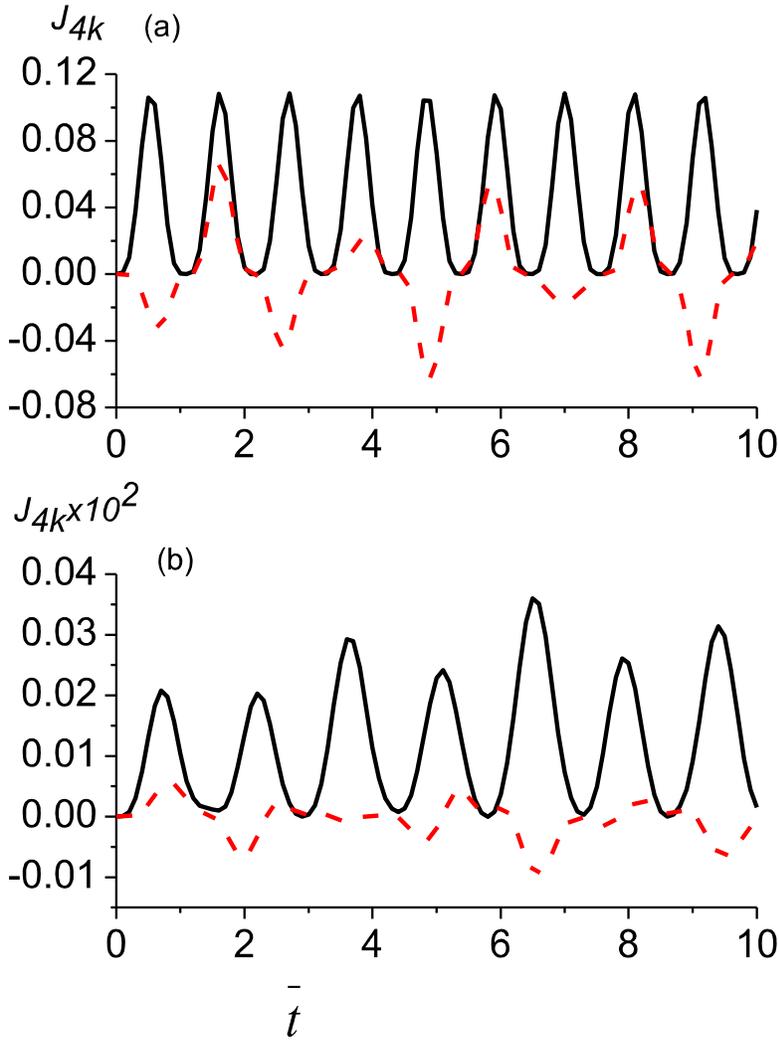

Fig. 3 The time dependence of the intensities of the MQ coherences of the fourth order in circles of four (a) and six (b) spins coupled by the DDI at low temperatures, $\beta = 10$. Black solid - $PS$ method, red dash - $QS$ method. The index "$k$" equals "$D$" and "$z$" for the results obtained by the $PS$ and $QS$ methods, respectively.

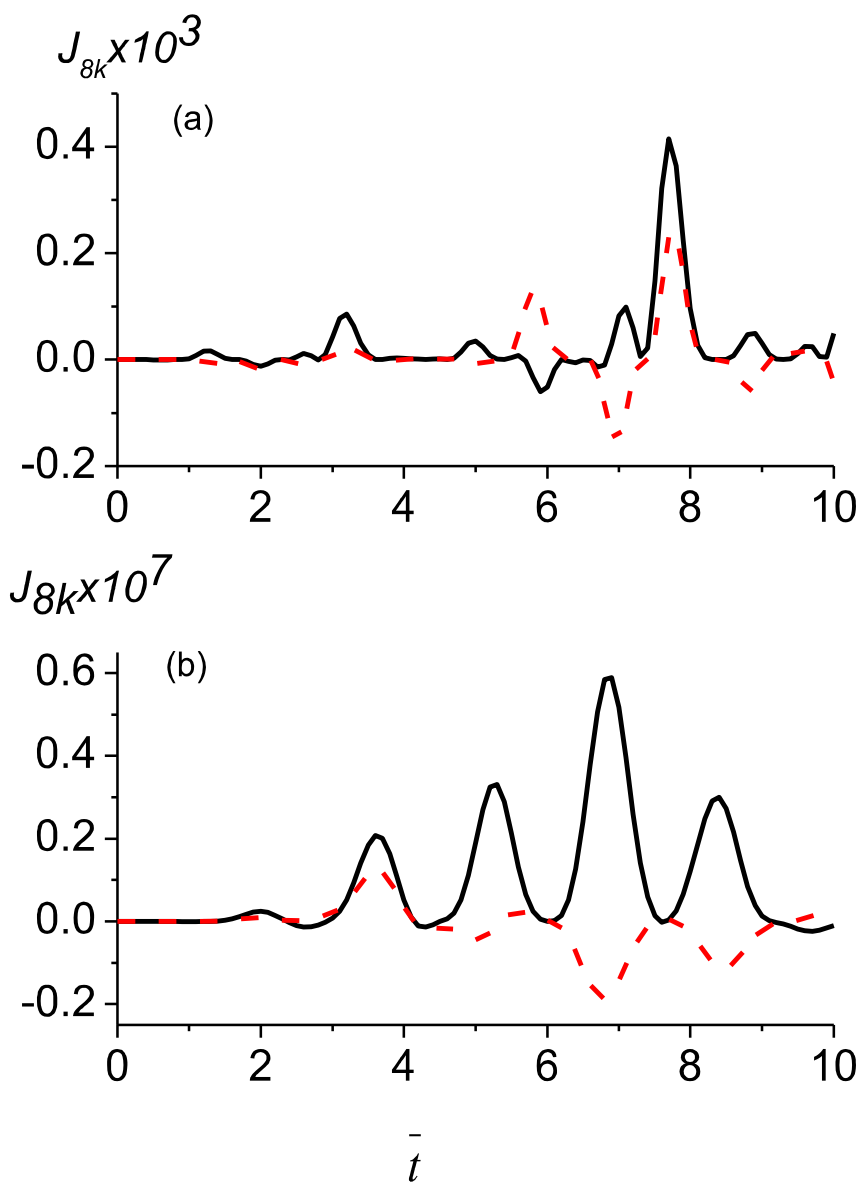

Fig. 4 The time dependence of the intensities of the MQ coherences of the eighth order in (a) a chain of eight spins and (b) cyclopentane molecules with ten spins (two pentagon cycles), $C_5H_{10}$. Black solid -- $PS$ method, red dash -- $QS$ method. The index " $k$ " equals " $D$ " and " $z$ " for the results obtained by the $PS$ and $QS$ methods, respectively.